\RequirePackage[hyphens]{url}
\documentclass[
 twocolumn,
nofootinbib,
 amsmath,amssymb,
 aps,
pra,
]{revtex4-2}

\usepackage{graphicx,amssymb,amsmath}
\usepackage{xfrac}
\usepackage{courier}
\usepackage{booktabs}
\usepackage[hyphens]{url}
\usepackage{comment}
\usepackage{lineno}
\usepackage{cancel}
\usepackage[normalem]{ulem}
\usepackage{xcolor}

\usepackage[breaklinks=true]{hyperref}

\newcommand{\ulysse}[1]{{\color{black} #1}}

\begin{document}

\title{Breakdown of the circular law from long-range correlations}

\title{Spectral properties of non-Hermitian real random matrices with \\
long-range correlations}


\author{Ulysse Marquis$^{1,2}$}
\email{ulyssepierre.marquis@unitn.it}

\affiliation{$^1$ Fondazione Bruno Kessler, Via Sommarive 18, 38123 Povo (TN), Italy}
\affiliation{$^2$ Department of Mathematics, University of Trento, Via Sommarive 14, 38123 Povo (TN), Italy}

\begin{abstract}


We investigate the spectral properties of non-Hermitian real random matrices whose entries exhibit long-range correlations decaying as~$|r-r'|^{-\alpha}$. \ulysse{We find a progressive breakdown of the circular law, controlled by the decrease of~$\alpha$. In all cases, the radial eigenvalue density decreases away from the origin. At~$\alpha>1$, an effective radius, reminiscent of the circular law, is retrieved, while instead, for~$\alpha<1$, the eigenvalue distribution broadens with matrix size and its spectral radius grows like a power law, with exponents numerically close to the exponents controlling the magnitude of fluctuations in the extended central limit theorem. The case~$\alpha=1$ appears as a case with self-similar eigenvalue density, and slowly growing spectral radius. Long-range correlations also enhance clustering of real eigenvalues and slow the resorption of the Saturn effect. These results reveal a correlation-driven transition and suggest the emergence of a new universality class for correlated non-Hermitian random matrices. }

\end{abstract}

\maketitle

\section{Introduction}
Long-range correlations are ubiquitous across physical systems~\cite{Sahimi_1993, Sagues_2007,Cross_1993}, in the form of power-law decay of the two-point function. \ulysse{In statistical mechanics, they are a hallmark of criticality~\cite{Stanley_1987}. They also appear in statistical analysis of financial markets or seismic data, and when applying detrending techniques~\cite{Podobnik_2008, Drozdz_2025}. Often, such systems can be represented under matrix form -- for instance, when embedded on a discrete square lattice.} One may naturally wonder then what are the properties of these random matrices with intricate correlations, which are often studied through their spectrum, as it contains relevant properties for many applications, such as diffusion, connectivity, robustness and community structure~\cite{Chung_1997,Barrat_2008}. Random matrix theory~\cite{Bouchaud_2020} has developed a powerful toolbox to investigate laws of the spectra of matrices with random independent (and often identically distributed, denoted i.i.d) entries. It resulted in the celebrated (semi-)circular laws, Marcenko-Pastur distribution and intricate mappings to electrostatic problems through log-gases~\cite{Forrester_2010}. While these frameworks have demonstrated their analytical efficiency and robust universality, their extension to objects with non-trivial sparsity~\cite{Rogers_2009}, block structure~\cite{Patil_2024} and correlations~\cite{Aceituno_2019,Baron_2022}, remains challenging, with complex networks~\cite{dorogovtsev2003} being a paradigmatic example. \\

\begin{figure}[htbp!]
    \centering
    \includegraphics[width=1\linewidth]{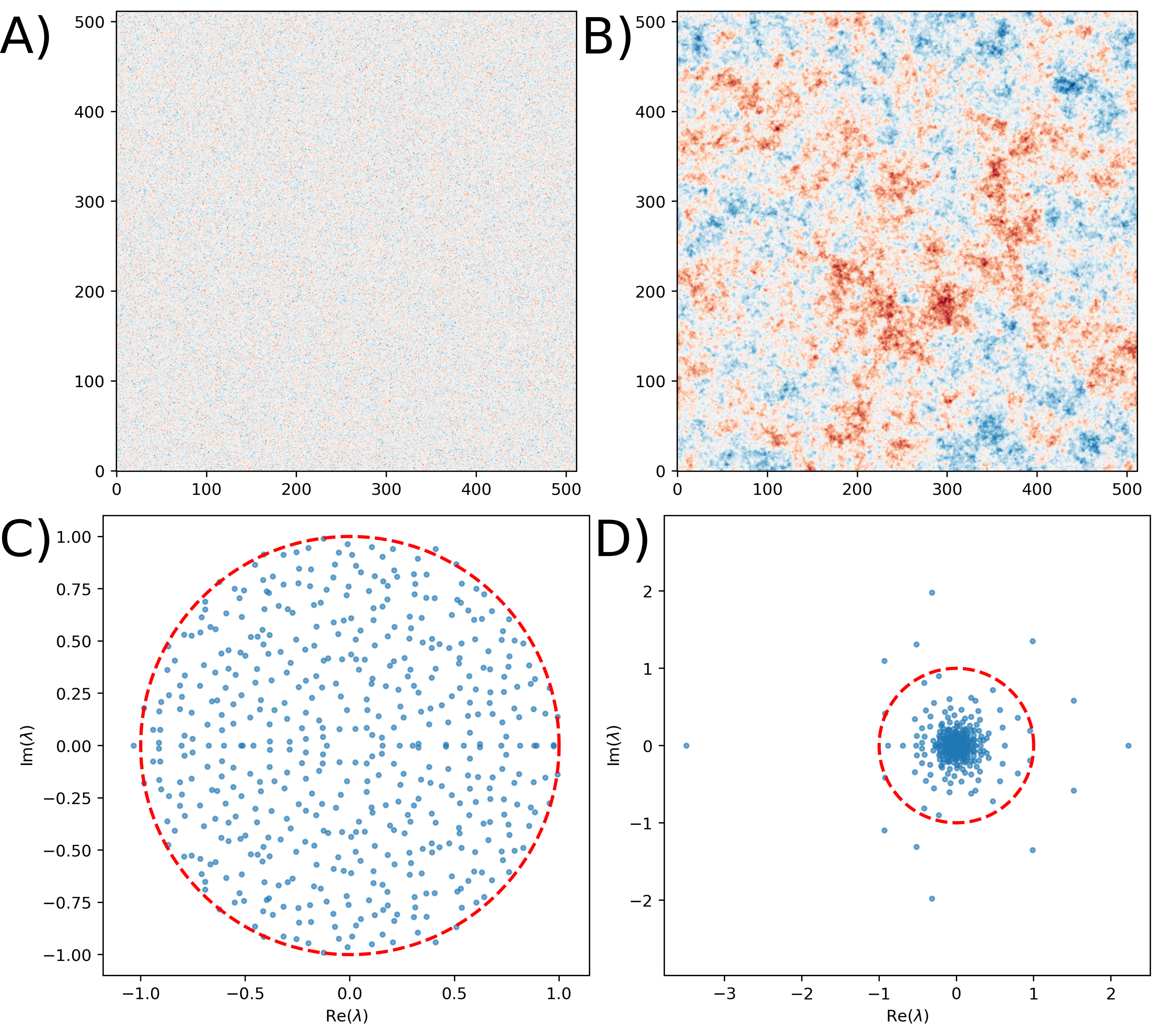}
    \caption{{\bf Ginibre against correlated matrix. } A) A~$512\times512$ Ginibre matrix : no large-scale structure emerges. B) Matrix of the same dimension with long-range correlations, controlled by~$\alpha=0.1$. \ulysse{Values in both matrices are displayed on the same color scales.} C) Spectrum of the Ginibre matrix : each eigenvalue corresponds to a point, the dotted circle corresponds to the asymptotic limiting boundary given by the circular law. Note that the density seems almost uniform, except around the real axis~$\Im\lambda = 0$, where eigenvalues are attracted by the line and deplete its surrounding : the `Saturn effect'. D) Spectrum of the matrix shown in B), the density is not uniform, but steeply decreasing with distance to~$\lambda=0$, and its eigenvalues are not contained in the red circle.}
    \label{fig:illustration}
\end{figure}

\begin{figure*}
    \centering
    \includegraphics[width=0.8\linewidth]{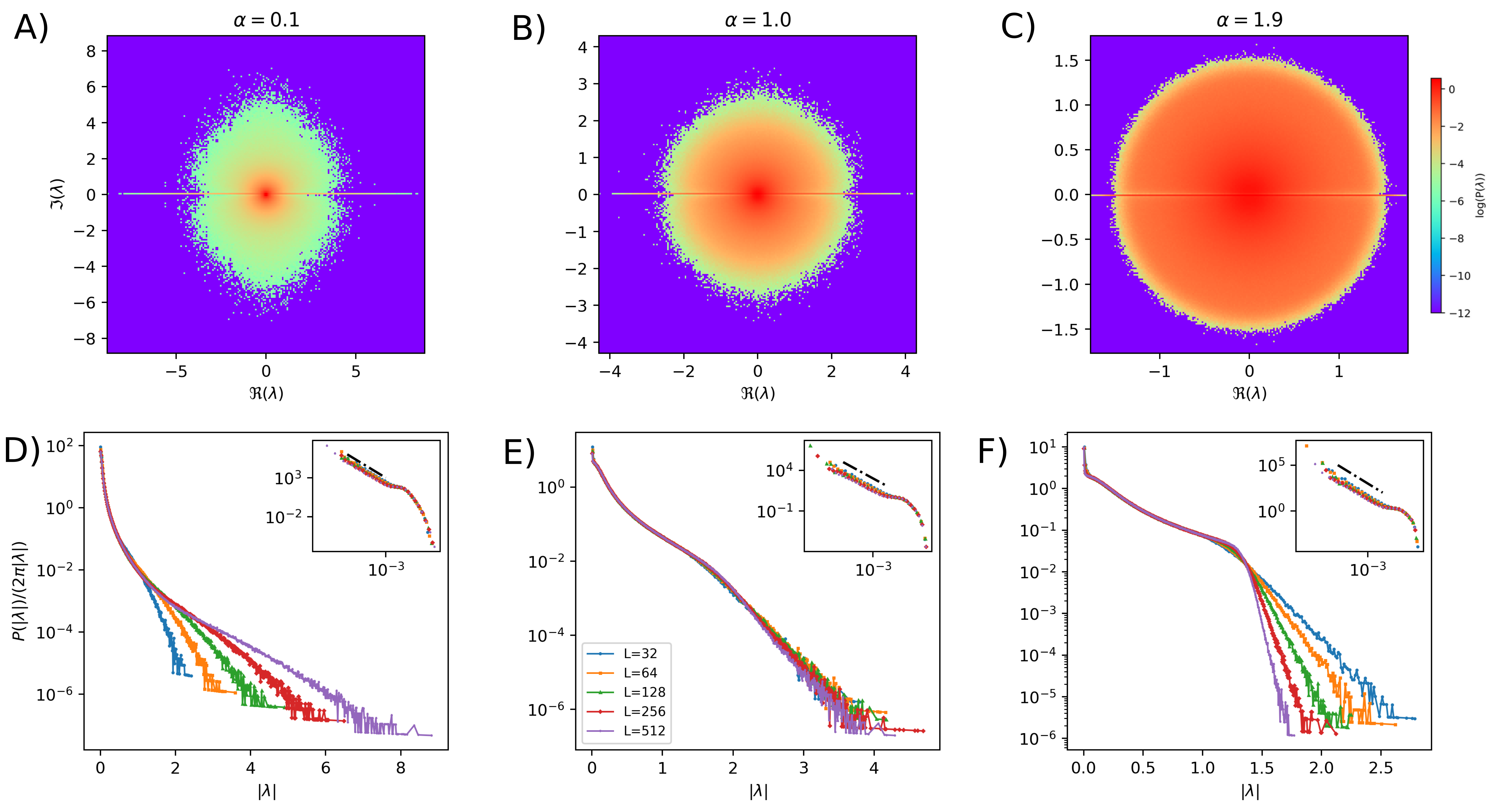}
    \caption{Two-dimensional histogram of eigenvalues of~$512\times512$ matrices for~$\alpha=0.1$ (A),~$\alpha=1$ (B) and~$\alpha=1.9$ (C). \ulysse{The histograms were computed by computing the spectra of~$5 \cdot 10^4$ matrices.} Note that the spectra exist far beyond the limit imposed by the circle law, in all cases, as well as the scale of the real eigenvalues when~$\alpha\leq1$.
    D)E)F) Empirical distribution of the modulus of the eigenvalues~$\mathrm{P}(|\lambda|)$ divided by~$2\pi|\lambda|$, for each value of~$\alpha$ and for~$L$ varying between~$L=32$ and~$L=512$. The baseline is the limit circle law~$\mathrm{P}(|\lambda|) / 2 \pi |\lambda| \sim \mathbf{1}_{\lambda<1}$. At~$\alpha<1$, the tail of the distribution broadens, while it narrows for~$\alpha>1$. The case~$\alpha=1$ appears as a self-similar case, as the densities collapse far from~$|\lambda|=0$. Spectra were computed by sampling~$5 \cdot 10^4$ matrices. \ulysse{Insets: zoom on the~$0 < \lambda < 1$ region. A histogram with log-spaced bins is displayed in log-log scale. The lines represent a power law of exponent~$-1$.}}
    \label{fig:density}
\end{figure*}

The goal of this paper is to investigate how the spectrum of matrices, representing~$2D$ lattices of linear size~$L$, with Gaussian entries
\begin{equation} \label{eq:magnitude}
    \mathrm{P}(M_{ij}) \propto \exp \left( -\frac{L M_{ij}^2}{2 }\right)
\end{equation}
is affected by long-range correlations
\begin{equation} \label{eq:correlations}
    \langle M_r M_{r'} \rangle \sim |r-r'|^{-\alpha} \,,
\end{equation}
where~$r=(i,j)$ ($r'=(i',j')$) and~$0<\alpha<2$, with respect to the i.i.d case, called the Ginibre ensemble~\cite{ginibre}\footnote{Here, the correlations are neither sparse (as in e.g.~\cite{Aceituno_2019}) nor constant across a small number of groups of pairs of entries as in~\cite{Baron_2022}. Instead, they cover a large range of values dependent on matrix entries distance.}. This exercise is typical when evaluating the range of validity statistics theorems: the universality of the central limit theorem (CLT) is known to hold for small range correlations~\cite{bouchaud1990}. Instead, for centered stationary sequences~$(X_1,\ldots,X_n,\ldots)$ with covariance~$\langle X_i X_{i+n} \rangle$ decaying slower than~$1/n$, such as~$1/n^\alpha$ with~$\alpha<1$, the order of the fluctuations of 
\begin{equation}
    S_n = \sum_{i=1}^n X_i
\end{equation}
is~$n^{1-\alpha/2}$ (or~$\sqrt{n \log n}$ for~$\alpha=1$). The distribution of~$S_n$ might also be non-Gaussian. \\

The sampling of matrices following eq.~\ref{eq:magnitude} and~\ref{eq:correlations} turns out to be a non-trivial task when the system size becomes large as for sampling from \ulysse{such an ensemble} one needs to store covariance matrices with~$L^2 \times L^2$ elements. However,~\cite{Makse_1996} proposed an efficient technique which allows to sample very large matrices -- the computational bottleneck then becomes finding the eigenvalues of non-Hermitian matrices, which amounts in practice to~$\mathcal{O}(L^3)$~\cite{pan1999}. This technique can be summarized in the following way: consider an i.i.d Gaussian sequence~$(u_r)$, the correlation function
\begin{equation} 
    C(\ell) = (1+\ell^2)^{-\alpha/2}\,,
\end{equation}
and its Fourier transform~$S(q)$ . Then, the inverse Fourier transform of the sequence~$\eta_q = \sqrt{S(q)} u_q$, where~$(u_q)$ are the Fourier transform coefficients of~$(u_r)$, yields the searched matrices~$(M_{ij})$. \ulysse{It is important to note that this convolution-based method imposes periodic boundary conditions.} An example for~$L=512$ and~$\alpha=0.1$ can be appreciated in Fig.~\ref{fig:illustration} and compared to a Ginibre realization. \ulysse{The panels (Fig.~\ref{fig:illustration}A,B)) display the matrix entries on a common color scale. By construction, the marginal distribution~$\mathrm{P}(M_{ij})$ is identical in both cases. The difference between the two panels lies instead in their spatial correlations: the correlated matrix exhibits long-range structure, with extended regions where neighboring entries take values in a similar range, whereas the Ginibre matrix is spatially uncorrelated. This clustering produces a stronger visual contrast in the correlated case, even though the underlying marginal distribution of entries is the same in both ensembles.}

\begin{figure*}[htbp!]
    \centering
    \includegraphics[width=0.8\linewidth]{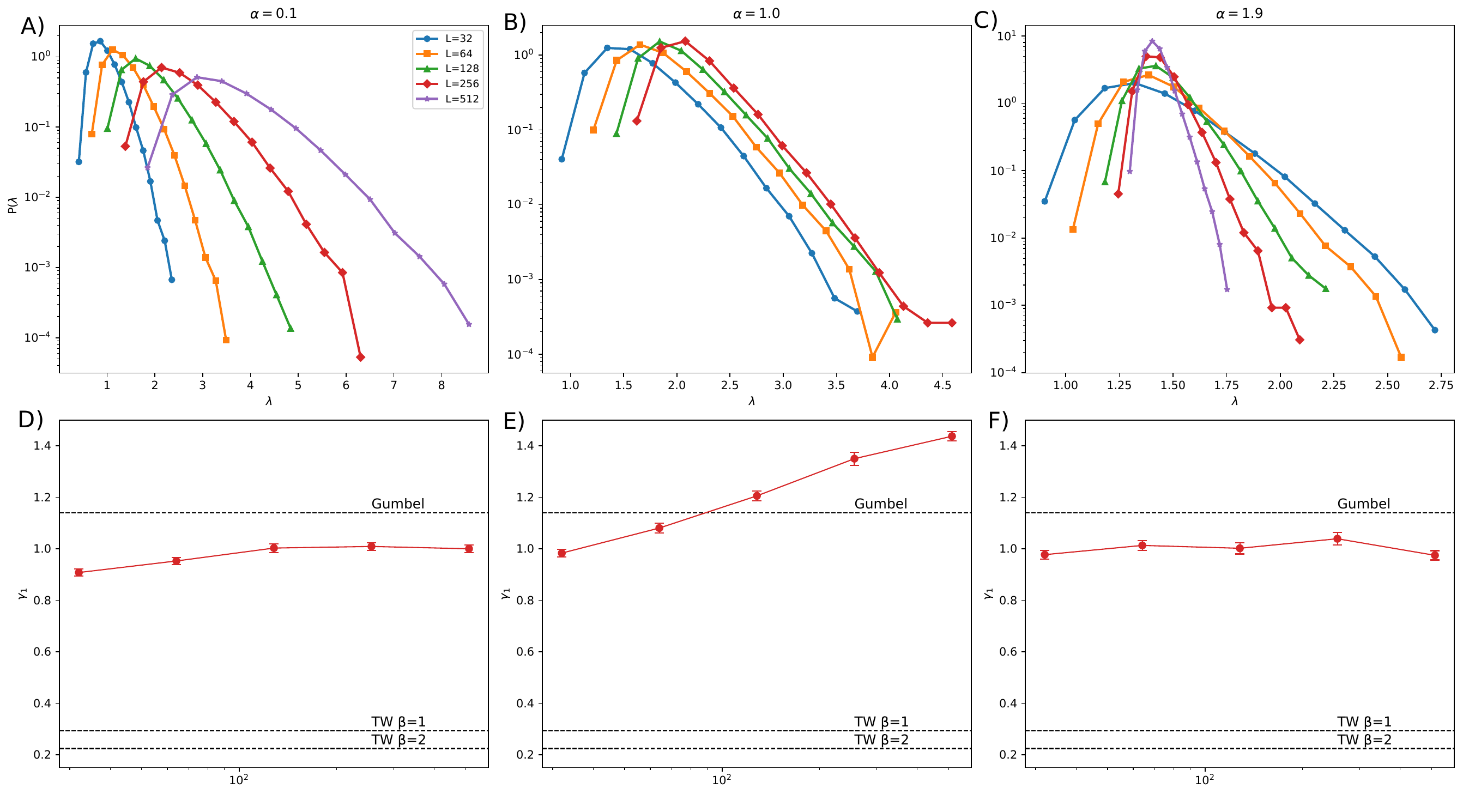}
    \caption{Distribution of the spectral radius~$\lambda_m$ when varying, for~$\alpha=0.1$ (A),~$\alpha=1$ (B) and~$\alpha=1.9$ (C). The position and scale of the distribution vary with~$L$ for~$\alpha\leq1$, while the distribution shrinks around a limit value ($r_e(\alpha=1.9) \approx 1.4$), corresponding to an effective radius for a distorted circle law, at rate~$L^{-1/2}$ when~$\alpha>1$. D)E)F) Skewness~$\gamma_1$ against matrix size~$L$ when varying~$\alpha$. Horizontal lines correspond to the skewness of baseline distributions: Gumbel ($\gamma_1 \approx 1.14$), Tracy-Widom for~$\beta=2$ ($\gamma_1\approx0.22$) and~$\beta=1$ ($\gamma_1\approx0.29$). The values of $\lambda_m$ were computed from~$5 \cdot 10^4$ realizations.}
    \label{fig:radius}
\end{figure*}

\section{Real Ginibre ensemble}

The real Ginibre ensemble was introduced in~\cite{ginibre} and occurred to be the most challenging ensemble -- analytically speaking -- among the `Gaussian' ones. It was found that when the entries have variance~$\frac{1}{L}$, the eigenvalues fill the complex plane \ulysse{according to} the circular law
\begin{equation}
    \rho(z) = \frac{1}{L} \sum_{i=1}^L\delta(z-\lambda_i) \xrightarrow[L \to \infty]{} \frac{1}{\pi}\mathbf{1}_{|z|<1}
\end{equation}
when~$L$ goes to infinity~\cite{tao2008}. The joint distribution of eigenvalues, conditioned on the number of eigenvalues, was computed in~\cite{lehmann1991}. It was found in particular that the probability to have~$L$ real eigenvalues was non-null and of order~$2^{-L(L-1)/4}$~\cite{edelman1997}, while the average number of real eigenvalues is~$\sqrt{\frac{2}{\pi} L}$ (to leading order)~\cite{Edelman1994}. The probability to find~$k$ real eigenvalues was computed in~\cite{kanzieper2005}. Correlation functions for real and complex eigenvalues were derived in~\cite{forrester2007,borodin2008} through the derivation of Pfaffian kernels. Extremal statistics, fundamental in the context of May's instability~\cite{May_1972}, were investigated in~\cite{Rider_2014,poplavskyi2016, baik2020,Cipolloni_2022,cipolloni2023}, in particular~\cite{Rider_2014} showed that the spectral radius converges to a Gumbel law. See also a recent review~\cite{byun2023} on the real Ginibre ensemble.

\section{Numerical investigations}

\subsection{Eigenvalue density}

Fig.~\ref{fig:illustration} shows the location of eigenvalues for one realization of the long-range correlated matrix ensemble, with~$\alpha=0.1$ and~$L=512$. Some features of the distribution immediately appear: a) the eigenvalues do not fill approximately uniformly the unit circle (displayed in red), but instead show a density decaying with the magnitude of the eigenvalues~$|\lambda|$; b) its edge is not \ulysse{sharply} decaying at~$|\lambda|=1$, instead several eigenvalues are located far from the boundary. This can be better understood by inspecting the eigenvalue density, shown in Fig.~\ref{fig:density}. Three behaviors emerge: at~$\alpha>1$, a bounded law with rescaled radius\footnote{\ulysse{The effective radius~$r_e$ is defined, when it exists, as the characteristic scale delimiting the bulk of the spectrum, i.e the radius beyond which the eigenvalue density sharply decays.}}~$r_e(\alpha)\approx 1.4$ at~$\alpha=1.9$ is recovered (see Fig.~\ref{fig:density} C),F)); instead, for~$\alpha<1$, the radial eigenvalue distribution is heterogeneous, with a marked peak at~$|\lambda|=0$ and \ulysse{broadens with increasing~$L$}. The case~$\alpha=1$ emerges as an essentially self-similar limiting case (see the collapse of the modulus density in Fig.~\ref{fig:density}E)). \ulysse{However, we find that for all~$0<\alpha<2$, in the vicinity of $\lambda \sim 0$, the number of eigenvalues at distance $r$ (more precisely, in an interval $[r, r+dr]$) is approximately constant (see insets of Fig.~\ref{fig:density}), whereas for the circular law it increases proportionally to $r$. This behavior indicates a strong accumulation of eigenvalues near the origin. In particular, since the number of eigenvalues per annulus is constant, the associated two-dimensional density scales as $r^{-1}$, revealing a singular peak at $\lambda = 0$.}

The distribution of~$\lambda$ does not tell the full picture as the distribution of eigenvalues in the complex plane are not isotropic: close to the real line~$\Im ( \lambda) = 0$, there is a discrepancy of eigenvalues and they are typically lying on the real line; furthermore, the range of the real eigenvalues typically overshoots the spectral radius -- this is called the~\textit{Saturn effect} and it vanishes when~$L \to \infty$ in the Ginibre case. It is clearly visible in Fig.~\ref{fig:density}A),B). For~$\alpha>1$, we retrieve a situation similar to the Ginibre case, for which the discrepancy disappears rapidly.

\begin{figure}
    \centering
    \includegraphics[width=1\linewidth]{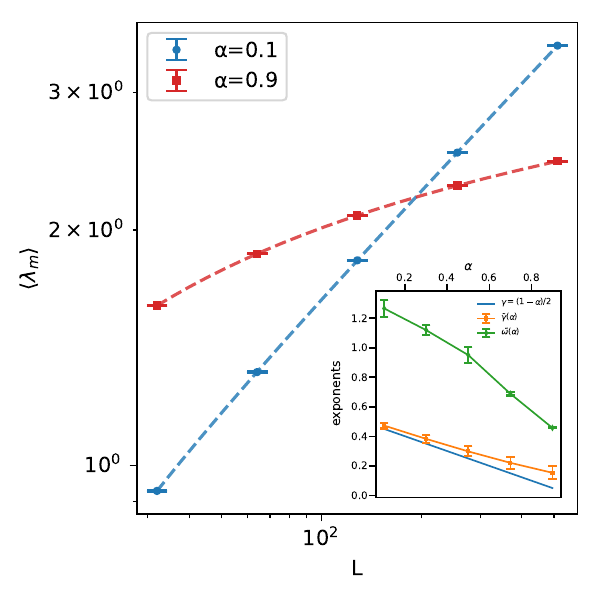}
    \caption{{\bf Scaling law for the spectral radius.} Main panel: measured average spectral radius~$\langle \lambda_m \rangle$ in function of system size for two values of~$\alpha$ given in the legend. The error bars represent the~$95\%$ confidence interval on the average. The dashed line represent the fit given by Eq.~\ref{eq:finitesize}. The averages are computed over~$5 \cdot 10^4$ realisations. Inset: measured scaling~$\tilde{\gamma}$ (orange points) and finite-size~$\tilde{\omega}$ (green points) exponents in function of the correlation exponent~$\alpha$. Blue line: extended CLT exponent~$\gamma=\frac{1-\alpha}{2}$. Error bars represent~$95\%$ confidence interval on the measured exponents.}
    \label{fig:exponents}
\end{figure}

\begin{figure*}
    \centering
    \includegraphics[width=0.8\linewidth]{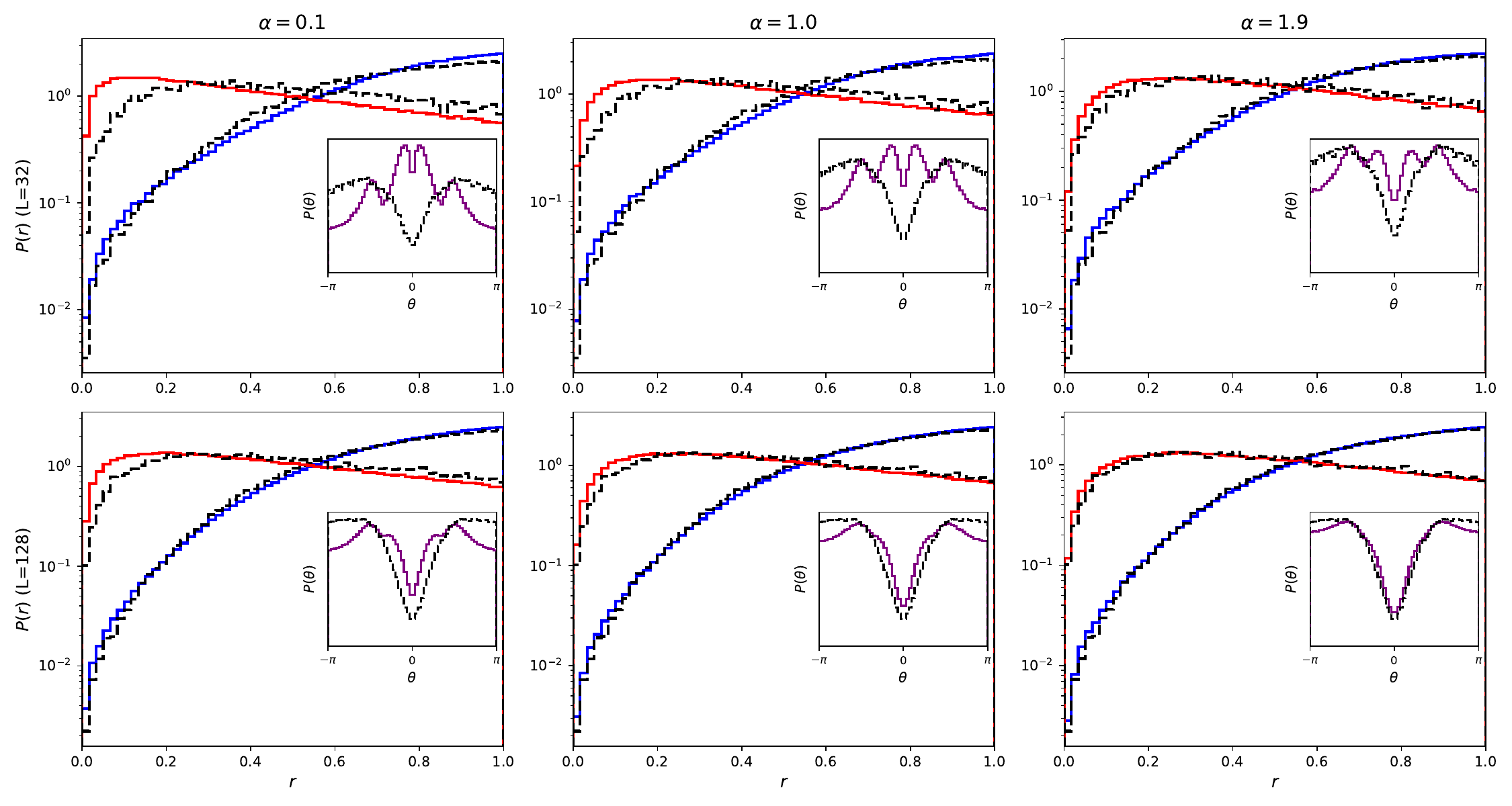}
    \caption{Eigenvalue spacing ratio distributions for~$L=32$ (top row) and~$L=128$ (bottom row) for various values of~$\alpha$. The main panels show the distributions of the magnitude of the ratios~$r = |z|$ of real eigenvalues (red curves) and complex eigenvalues (blue curves), compared to their Ginibre counterparts (black curves). The inset displays the argument~$\theta = \arg z$ distribution. Each curve was computed using the eigenvalue distribution from~$5 \cdot 10^4$ diagonalized matrices, and~$5 \cdot 10^3$ for the Ginibre ensembles.}
    \label{fig:ratio}
\end{figure*}

\subsection{Spectral radius}

A quantity of interest is the breadth of the distribution of eigenvalues,~$\lambda_m = \max |\lambda_i|$. It plays a particular role in controlling the equilibrium of large dynamical systems~$\dot{\mathbf{x}} = f(\mathbf{x})$ for which the Jacobian around an equilibrium can be written~$J=-d I+A$, where~$d$ is the self-regulation term while~$A$ describes the interactions and is often modeled as random. Then, the stability condition of such equilibrium can be written
\begin{equation} \label{eq:may}
    \lambda_m < d \,,
\end{equation}
where~$\lambda_m$ typically scales as~$\sqrt{L}$~\ulysse{in the case of uncorrelated matrices}. This is the so-called May's stability criterion which inspired a large body of works in ecology~\cite{May_1972, Fyodorov_2016, Allesina_2015} and more recently has found applications in the study of large economies~\cite{Moran_2019}.

The phenomenology observed for the tails of the distribution of~$\lambda$ in Fig.~\ref{fig:density} can be better understood by looking at the distribution of~$\lambda_m$, shown in Fig.~\ref{fig:radius}. For~$\alpha>1$, the scale of the distribution decays with~$L$, as illustrated in Fig.~\ref{fig:radius}C). More precisely, the standard deviation~$\sigma(\lambda_m)$ decays following the scaling law~$\sigma(\lambda_m) \sim L^{-1/2}$. For~$\alpha \leq 1$, the situation changes as the position and the scale of the distribution of~$\lambda_m$ varies with~$L$ (see Fig.~\ref{fig:radius}A),B)). \ulysse{Plotting the average spectral radii as a function of the system size, as shown in Fig.~\ref{fig:exponents} (main panel), we find that they follow scaling laws
\begin{align}
    \langle \lambda_m \rangle \sim L^\gamma \,.
\end{align}
Measuring directly these exponents ($\tilde{\gamma}$, orange points in the inset of Fig~\ref{fig:exponents}), we find a good agreement with the exponent predicted for the fluctuations in the extended CLT (blue line)
\begin{align}
    \gamma = \frac{1-\alpha}{2} \,.
\end{align}
We find that the difference is attributable to finite-size effects, and when fixing~$\gamma$, a scaling form 
\begin{align} \label{eq:finitesize}
    \langle \lambda_m \rangle \sim L^\gamma (1- k \, L^{-\omega}) \,,
\end{align}
is found to describe well the observed variations. Examples of fits are shown in the main panel of Fig.~\ref{fig:exponents} (dashed lines) and the measured finite size exponents~$\tilde{\omega}$ are shown in the inset (green points).}

In the case~$\alpha=1$, the position of the distribution instead displays sub-power-law behavior
\begin{align}
    \langle \lambda_m  \rangle \sim \sqrt{\log L}
\end{align}
also consistent with the CLT prediction. Beyond its first two moments, higher order moments and the full distribution of the fluctuations of~$\lambda_m$ are of interest. The skewness is shown as a function of matrix size in Fig.~\ref{fig:radius}. \ulysse{In the cases~$\alpha \neq 1$, while it shows clear departure from the Tracy-Widom ensembles, the skewness values remain close, but distinct, to the Gumbel skewness, expected asymptotically in the Ginibre case. Instead, when~$\alpha=1$ the skewness increases with~$L$ and largely overshoots these values, pointing towards different extreme statistics phenomenology.}

\subsection{Eigenvalue spacing ratio}

Beyond density, level spacings are typically of interest as they characterize for instance quantum chaos (see~\cite{Sa_2020} and references therein). While the distribution of distances between consecutive levels~$s = \lambda_{i+1}-\lambda_i$ suffices for Hermitian ensembles, there is no \ulysse{natural} ordering in the complex plane and one needs to employ other means. The ratio
\begin{align}
    z = \frac{\lambda_k - \lambda^{\text{NN}}_k}{\lambda_k - \lambda^{\text{NNN}}_k} \,,
\end{align}
where~$\lambda^{\text{(N)NN}}$ is the (next-to)-nearest-neighbor eigenvalue, was introduced in~\cite{Sa_2020} as an observable for level spacing in the complex plane. Here, we compute separately the statistics of spacing for the real eigenvalues and strictly complex eigenvalues. In Fig.~\ref{fig:ratio} is displayed the density of the magnitude of ratios for the real eigenvalues (red curve), complex eigenvalues (blue curve) and of the arguments (purple curve, insets), and they are compared to equivalent Ginibre ensembles (black curves). \\

The mean spacing ratios $\langle r \rangle = \langle|z| \rangle$, quantifying tendency for eigenvalues to cluster ($r \sim 0$) or show level repulsion ($r \sim 1$), reveal distinct behaviors for the complex and real sectors. In the complex case, the density are consistently similar between the correlated matrices ensemble and Ginibre, and also when varying~$\alpha$ and~$L$. The average is about~$\langle r \rangle \approx 0.71 - 0.74$. In contrast, the real sector shows systematically~$r$ smaller ($\langle r \rangle = 0.42-0.49$) than for the Ginibre ensemble ($\langle r \rangle \approx 0.49-0.50$), sign of enhanced clustering of eigenvalues on the real line. This disparity decreases as~$\alpha \to 2$ and also when~$L$ increases. Instead, the distribution argument~$\theta = \arg z$ varies significantly when varying~$\alpha$ and~$L$.~$\theta$ reflects preferred alignment directions of neighboring eigenvalues, and quantifies deviations from rotationally symmetric distributions. For the real Ginibre ensemble, the mean $\langle \cos\theta \rangle$ is slightly negative, with a decreasing trend, from~$\langle \cos \theta \rangle\approx- 0.11$ at~$L=32$ to~$\approx -0.20$ at~$L=256$. Instead, for~$\alpha=0.1$,~$\langle \cos \theta \rangle \approx 0.21 > 0$ at~$L=32$ before decreasing to~$\approx -0.11$ at~$L=256$. Similar patterns, of lower amplitudes, can be found for ensembles with higher values of~$\alpha$. However, at any~$\alpha$, the argument of the ratios converge to the Ginibre distribution.

\begin{figure}
    \centering
    \includegraphics[width=1\linewidth]{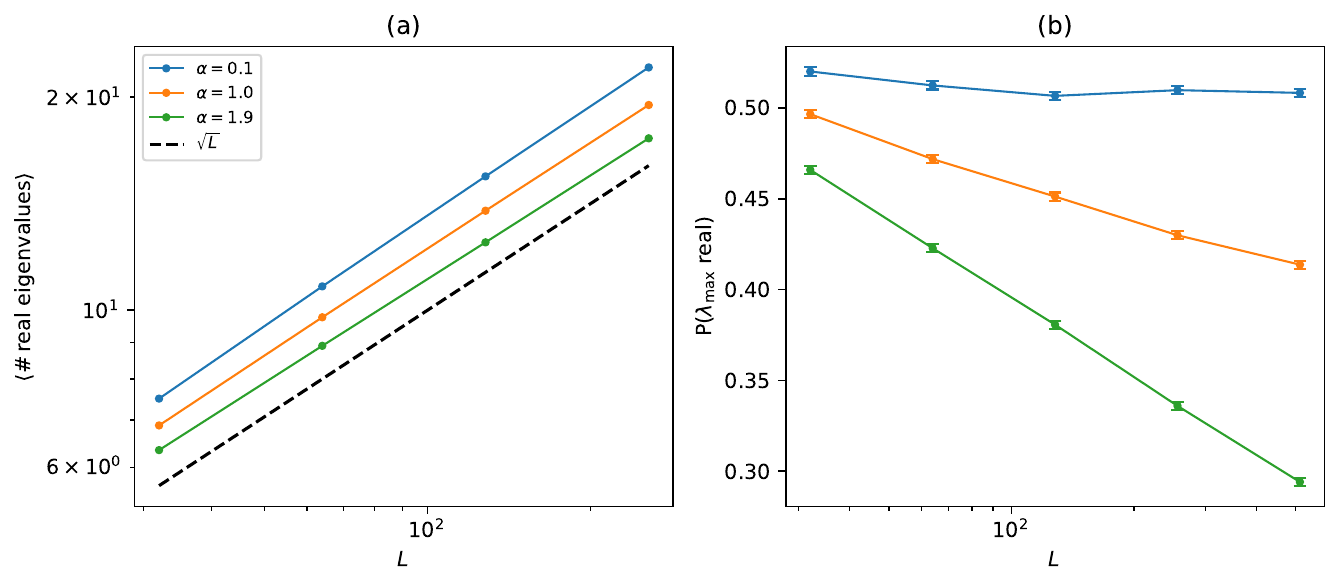}
    \caption{a) Expected number of real eigenvalues as a function of~$L$. Black dashed line:~$\sqrt{L}$. b) Empirical probability for the leading eigenvalue to be real against matrix size~$L$.}
    \label{fig:real}
\end{figure}

\subsection{Real eigenvalues}

Eigenvalues of random matrices are roots of the associated characteristic polynomials, either real or conjugated complex. Regarding the expected number of real roots of random polynomials, essentially two regimes appear, depending on the variance of the coefficients. When the coefficients are i.i.d standard normal, then the number of real eigenvalues grows like the logarithm of the degree; when the $i$-th coefficient have variance~$\binom{n}{i}$, it grows like the square root of the degree~\cite{edelman1995zerosrandompolynomialreal}. This is in particular the case of eigenvalues of random matrices. In the case of long-range correlated matrices, the number of real eigenvalues as a function of~$L$ is shown in Fig.\ref{fig:real}. While their number remains proportional to~$\sqrt{L}$, the prefactor varies from the Ginibre value~$\sqrt{2/\pi}$, and increases when~$\alpha$ decreases. \\
In the case of the Ginibre ensemble, the `Saturn effect' is quickly resorbed and the probability that the leading eigenvalue is real decays with matrix size~\cite{Rider_2014} (the speed of this decay is not well yet understood to the best of our knowledge). Given Fig.~\ref{fig:density}, where the breadth of the real eigenvalues is the same as the complex eigenvalues' at large~$\alpha$, and not for~$\alpha \leq 1$, we probe the probability that the leading eigenvalue is real, as shown in Fig.~\ref{fig:real}. While it decreases notably for~$\alpha>1$, the decay rate is lower for~$\alpha=1$ and even more for~$\alpha<1$, with more than half of the leading eigenvalues being real up to~$L=512$. This observation raises questions about the appearance of a novel scale for real eigenvalues when there are strong long-range correlations: despite their small proportion~$\mathcal{O}(1/\sqrt{L})$, they dominate the high-end of the spectrum. 

\section{Discussion}

We have analyzed numerically the random matrix ensemble defined by Eq.~\ref{eq:magnitude} and~\ref{eq:correlations}, and evidenced deviations from the Ginibre ensemble, focusing on particular on a) the eigenvalue density b) the spectral radius c) the eigenvalue spacing d) the real eigenvalue properties. 
There are serious differences with the Ginibre phenomenology. \ulysse{The most salient one is the scaling law of the spectral radius when~$\alpha <1$, with 
exponents numerically coinciding with those controlling the magnitude of fluctuations in the extended central limit theorem~\cite{Majumdar_2020}. Given the numerical evidence, we conjecture the correspondance between these two scaling exponents -- establishing a formal mapping would yield great insight in the analytical understanding of correlated random matrices. 

Such a variation is relevant in the context of the study of large dynamical systems. Two important applications are (i) the stability of large ecosystems, as mentioned above with May's instability. Given the stability condition~(\ref{eq:may}), the correct scaling law for the spectral radius reads
\begin{align}
    \lambda_m \sim L^{\frac{1}{2}+\gamma} \,,
\end{align}
a spectral radius growing faster than the na\"ive May prediction~$\lambda_m \sim L^{\frac{1}{2}}$. Hence, systems with correlations such as those described by Eq.~\ref{eq:correlations} could exhibit reduced stability. (ii) Regarding the study of neural networks, spectral radius of random matrices is relevant for understanding chaotic behaviors~\cite{Sompolinksy_1988} and also for confronting the problems of \emph{exploding} or \emph{vanishing} gradients in the training of recurrent networks~\cite{Pascanu_2013}.} 

We hope that this investigation will stir analytical interest. A promising direction comes from works on cyclic correlations~\cite{Aceituno_2019}: from the Green function~$G$, one can derive the complex eigenvalue density~\cite{Rogers_2010}. The difficulty lies in computing the ensemble average of the Green function~$\overline{G}$. For cyclic correlations, writing~$G$ as a geometric series and using a first-order development yields precise predictions of the hypotrochoidic laws found in this case. Here, correlations appear at all orders. Estimating~$\overline{G}$ in this case is thus an open challenge.

\textit{Acknowledgments ---} The author is indebted to Andrea Legramandi for illuminating discussions and thanks the anonymous referees for their useful suggestions.

\end{document}